\documentclass[%
preprint,
preprintnumbers,
 amsmath,amssymb,
 aps,
 showpacs,
 jcp,
floatfix,
]{revtex4-1}

\usepackage{graphicx}% Include figure files
\usepackage{dcolumn}% Align table columns on decimal point
\usepackage{bm}% bold math
\usepackage[version=3]{mhchem}

\usepackage{natbib}
\usepackage{epstopdf}
\usepackage{mathtools}
\usepackage[font=footnotesize,labelfont=bf]{caption}
\graphicspath{{./figures/}}

\newcommand{\sprime}{{s^\prime}}
\newcommand{\Ca}{\ce{Ca^2+} }
\newcommand{\PO}{\ce{PO4^3-}}
\newcommand{\dd}{\mathrm{d}}
\newcommand{\shat}{\hat{s}}
\newcommand{\that}{\hat{t}}

\begin{document}
\title{Regulatory inhibition of biological tissue mineralization by calcium phosphate through post-nucleation shielding by fetuin-A}% Force line breaks with \\

\author{Joshua C. Chang}\email{joshchang@ucla.edu}
 \affiliation{Clinical Center, National Institutes of Health, Bethesda MD 20812}
 \affiliation{Mathematical Biosciences Institute, The Ohio State University, Columbus OH 43210}
\author{Robert M. Miura}
 \email{miura@njit.edu}
\affiliation{%
Department of Mathematical Sciences, New Jersey Institute of Technology, Newark NJ 07102
}%

\begin{abstract}

In vertebrates, insufficient availability of calcium and inorganic phosphate ions in extracellular fluids leads to loss of bone density and neuronal hyper-excitability. To counteract this problem, calcium ions are usually present at high concentrations throughout bodily fluids -- at concentrations exceeding the saturation point. This condition leads to the opposite situation where unwanted mineral sedimentation may occur. Remarkably, ectopic or out-of-place sedimentation into soft tissues is rare, in spite of the thermodynamic driving factors. This fortunate fact is due to the presence of auto-regulatory proteins that are found in abundance in bodily fluids. Yet, many important inflammatory disorders such as atherosclerosis and osteoarthritis are associated with this undesired calcification. Hence, it is important to gain an understanding of the regulatory process and the conditions under which it can go awry. In this manuscript, we extend mean-field continuum classical nucleation theory of the growth of clusters to encompass surface shielding. We use this formulation to study the regulation of sedimentation of calcium phosphate salts in biological tissues through the mechanism of post-nuclear shielding of nascent mineral particles by binding proteins. We develop a mathematical description of this phenomenon using a countable system of hyperbolic partial differential equations. A critical concentration of regulatory protein is identified as a function of the physical parameters that describe the system.
\end{abstract}

\maketitle

\section{Introduction}

In biology, ionic calcium (\Ca) plays many diverse roles including acting as a secondary messenger
in  biochemical cascades and modulating
neuronal excitability \cite{lu2010extracellular}. Structurally,
 \Ca and inorganic phosphate ions  (\ce{PO4^-}) are also the main constituents of bones in
vertebrates. For this reason, it is important for organisms to obtain adequate amounts of 
calcium from the environment. 

In normal circumstances,
\Ca is plentiful throughout extracellular spaces and in the circulatory system, stably existing 
 at concentrations \emph{exceeding} the saturation point, whereby sedimentation is favored~\cite{holt2014mineralisation}. Even under the 
 dangerous condition of hypocalcemia, \Ca may still be supersaturated relative to the most thermodynamically
 stable phase of calcium phosphate, hydroxyapatite (HAP), which is the building block of teeth and bones.
 
Yet, while the deposition of calcium into bones is desirable, ectopic calcification into soft tissues is pathological and
either causes or exacerbates  a variety of inflammatory disorders including arteriosclerosis, heart disease, and arthritis~\cite{smith2012phosphorylated,bendon1996histopathology,brylka2013role}. 
So, the regulation of ectopic calcium sedimentation is important in maintaining the health of soft tissues.

When calcium-phosphate solutions are supersaturated, HAP is formed in a multi-step process traversing through several intermediate crystalline or psuedo-crystalline states. The first step in this process is thought to be formation of pre-nucleation clusters~\cite{dey2010role,gebauer2014pre,habraken2013ion}, which are
small calcium-phosphate complexes~\cite{wang2012posner}.  Although some debate exists~\cite{habraken2013ion,gebauer2014pre,xie2014tracking}, these small complexes are thought to be \emph{Posner's clusters} (PC), with composition \ce{Ca9(PO4)6}~\cite{yin2003biological}.  A combination of experimental and theoretical analyses have confirmed the stability of PCs~\cite{treboux2000existence,treboux2000symmetry}, their presence in physiological solutions~\cite{yin2003biological}, and their consistency with the unit-cell structure of calcium-phosphate precipitates~\cite{dey2010role,du2013structure}.
For the purposes of this manuscript, we will assume that PCs are the fundamental building blocks of  larger-scale calcium-phosphate clusters and refer to them as monomers.

 These  monomers aggregate whereby they nucleate into amorphous spherical post-nucleation clusters composed  of amorphous calcium phosphate (ACP)~\cite{boskey1973conversion,wang2008calcium,dorozhkin2010amorphous,dorozhkin2011calcium}, having a calcium to phosphate ratio of approximately three to two~\cite{jiang2015amorphous}. As long as supersaturation persists,  and in the absence of regulation,  ACP clusters continue to grow by absorbing additional monomers into their structure. When ACP clusters become sufficiently large, they sediment into the tissue while simultaneously undergoing several phase transitions before eventually transforming into HAP.

This situation is seemingly incompatible with life as the persistent supersaturation of calcium and phosphate in biological fluids
dictates unyielding sedimentation, at least in the absence of regulatory inhibition. Fortunately
a regulatory mechanism does exist.
 The main machinery for preventing calcium phosphate sedimentation is the 
plasma protein fetuin-A (FA)~\cite{price2003inhibition,herrmann2012fetuin,brylka2013role,heiss2010fetuin,jahnen2011fetuin}. FA is an acidic protein that is found abundantly in blood as well as throughout
all extracellular compartments. Maintenance of adequate levels of FA protein has been shown to be necessary for inhibition of calcification in relation to many disorders~\cite{westenfeld2009fetuin,smith2013serum}. 

FA interacts with the calcium phosphate mineralization process in several
different ways. It can directly bind calcium, with each molecule able to weakly and reversibly bind
approximately 12--15 ions~\cite{suzuki1994calcium}. Yet, this direct binding of calcium cannot be the main regulatory mechanism
of FA as it would effectively reduce the supersaturation. There is also evidence that FA binds to pre-nucleation clusters~\cite{heiss2010fetuin}, a possibility that has been analyzed~\cite{wang2012theoretical}, although somewhat contradictory evidence has also shown that the presence of FA does not affect the rate of nucleation of calcium phosphate clusters~\cite{rochette2009shielding}. 
Primarily, FA binds strongly to post-nuclear calcium-rich calcium-phosphate clusters, shielding them from further growth and imparting upon them enhanced colloidal stability so that they do not sediment.

In this manuscript, we adapt mean-field classical nucleation theory (CNT) to look at the inhibition of mineral cluster growth by FA. We provide a quantitative description of the overall regulatory process and examine conditions necessary for stability.

\section{Quantitative methods}

The problem of understanding the combined process of mineralization and FA-induced inhibition is an example of a \emph{nucleation} problem. Our approach to this problem is to use ideas from mean-field \emph{classical nucleation theory} (CNT). In particular, we utilize a continuum approximation to the kinetic theory whereby we frame our problem using a series of serially-coupled partial differential equations (PDE).  It is notable, however,  that theoretical treatments of the inherently high-dimensional stochastic problem of nucleation and aggregation also exist~\cite{marcus1968stochastic,dorsogna2013combinatoric,d2012stochastic}. The mean-field theoretic CNT approach is ultimately motivated by the behavior of such stochastic treatments.

Our overarching goal in this section is to understand the kinetics of the concentration profile for mineral clusters as a function of size and interactions with FA. To this end, in Section~\ref{sec:nucleation}, we first solve for the nucleation rate (formation rate of critically-sized clusters), which provides a boundary condition for our PDE problem.
Then, in Section~\ref{sec:growth}, we derive an effective growth rate ($v_n(s))$ for the mineral portion of formed mineral and protein-mineral clusters as a function of their fixed mineral surface area $s$ and number of bound FA proteins $n$. Then, in Section~\ref{sec:shielding}, for a fixed cluster configuration, we derive the shielding rate under the assumption that it is governed by diffusion-limited kinetics. Finally, in Section~\ref{sec:continuum}, we tie together the various components of our theory (nucleation, growth, shielding) into an overarching continuum model.

For the reader's convenience, we have compiled a list of the mathematical symbols that we use throughout this manuscript into Table~\ref{tab:symbols}.

\subsection{Nucleation}
\label{sec:nucleation}

\begingroup
\squeezetable
\begin{table}
\begin{tabular}{lp{0.4\textwidth}}
Symbol & Description \\ \hline
$s$ & Surface area of mineral phase of cluster\\
$m$ &Number of mineral monomers in a cluster, where \\
&  monomer refers to Posner's cluster \ce{Ca9(PO4)6}. \\
$r$ & Radius \\ 
$V$ & Volume \\
$n$ & Number of shielding proteins attached to surface of cluster \\
$\gamma$ & Interfacial free energy per unit surface area  in units $k_BT$ per squared meter  \\
$f$ & Geometric correction for surface free energy for non-spherical pre-critical states \\
$\rho_\infty$ & Concentration of mineral monomers \\
$\rho_s$ & Saturation concentration\\
$\Delta\mu$ & $-\log(\rho_\infty/\rho_s)$, Chemical free energy per mineral monomer in units $k_BT$ \\
$\Delta G$ & Gibbs free energy  in units $k_BT$ \\
$c_n(s,t)$  & Concentration of mineral clusters \\ &of surface area $s$ shielded by $n$ FA monomers\\
$\sprime$ & Shielded surface area\\
$s_n$ &Amount of surface shielded ($\sprime$) when $n$ \\ & FA monomers are bound\\
$\delta s_n$ & $s_{n}-s_{n-1}$ for $n\geq 1$ \\
$s_*,m_*$ & Critical cluster size at nucleation \\
$s_p, m_p$ & Critical cluster surface area and monomer number at sedimentation \\
$D$ & Diffusivity of mineral monomers \\
$D_{\textrm{FA}}$  & Diffusivity of FA monomers \\
$k_-$ & Dissociation rate per unit surface area for mineral \\
$\phi_\infty$ & Concentration of FA monomers \\
$\bar{r},\bar{s},\bar{v}$ & Mineral monomer radius, surface area, volume \\
$\omega$ & $D\sqrt{4\pi}\rho_s/k_-$ \\
$\alpha$ & $8\bar{v}D\pi \rho_\infty$ \\
$\beta$ & $-\Delta\mu \sqrt{s_*}/f $ \\
$\varepsilon$ & Thickness of shielding layer (of FA protein) \\
$\lambda$ & $\sqrt{4\pi}D_{\textrm{FA}}\phi_\infty$  
\end{tabular}
\caption{List of mathematical symbols used in the manuscript for easy reference}
\label{tab:symbols}
\end{table}
\endgroup

  CNT explains the emergence and evolution of colloidal phases in solutions through the development of a simple thermodynamical picture. The key element of CNT is the assumption that the emergence of a new phase carries an energetic cost due to the creation of an interfacial surface.
  
\begin{figure}
\includegraphics[width=\columnwidth]{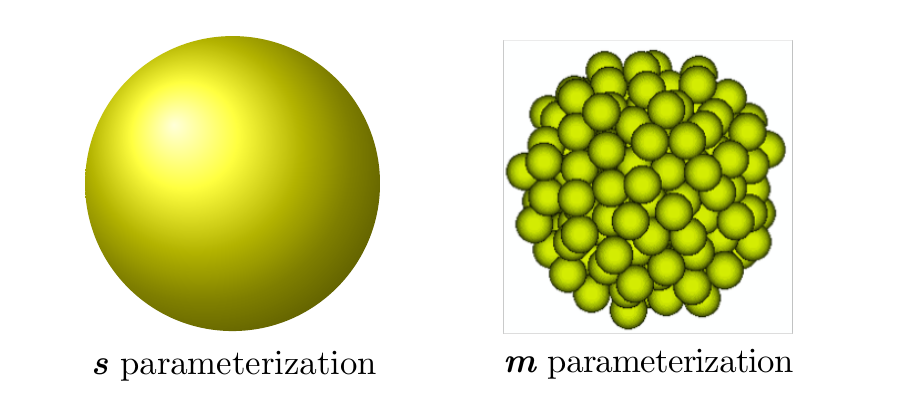}
\caption{\textbf{Cluster size parameterizations.} Surface area ($s$) and monomer count ($m$) parameterizations are used interchangeably for expressing the size of mineral clusters. Clusters are assumed to be spherical and composed of an integer number $m$ of monomers (Posner's clusters). The cluster as a whole has surface area $s$. To denote the conversion between these two parameterizations we use functions $m(s):s \to m$ and $s(m):m\to s$. }
\label{fig:parameterizations}
\end{figure}

  Consider a mineral particle consisting of an integer number $m$ mineral subunits, hereby termed an ``$m$-cluster.'' We say that this particle
has volume $V=m\bar{v}$, where $\bar{v}$ is the effective volume of each subunit ``monomer.'' Assuming that this particle is spherical, it has surface area $s = (36\bar{v}^2\pi)^{1/3}m^{2/3}$, and radius $r=\sqrt{s/4\pi}$. In this manuscript, we use both $m$ and $s$ to parameterize the size of mineral clusters (see Fig.~\ref{fig:parameterizations}). We also will refer to the conversion between these two parameterizations as $m(s)$ and $s(m)$.

\begin{figure}
\includegraphics[width=\columnwidth]{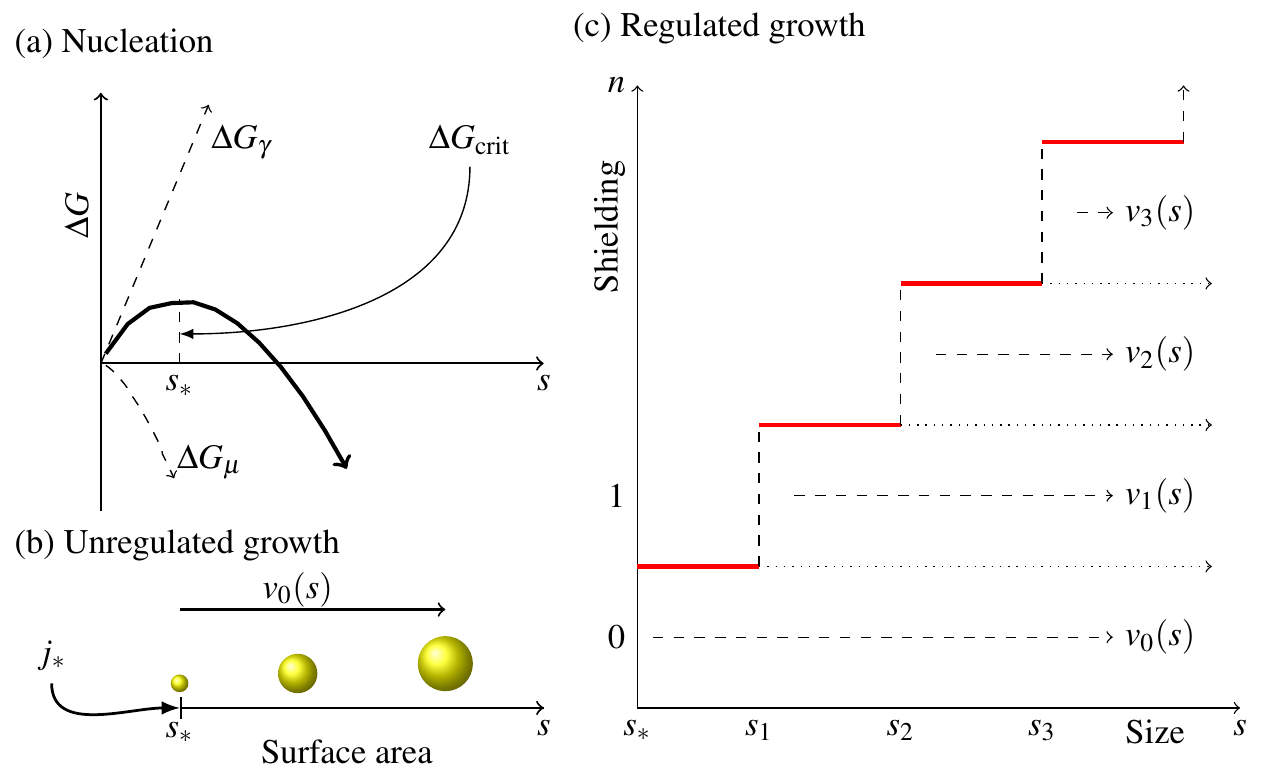}\vspace{-1em}
\caption{\textbf{Nucleation, growth, and shielding.} (a) \textbf{Classical nucleation} of an initial critically-sized spherical mineral cluster of surface area $s_*$ ($\approx3\textrm{nm}^2$). The activation energy $\Delta G_{\textrm{crit}}$ (units of $k_BT$) corresponds to the energy of the cluster of critical size (of units surface area).  A flux $j_*$ of nucleating particles is generated by the system under the condition of supersaturation. \textbf{(b) Unregulated growth.}  Upon nucleation, particles grow uncontrollably at a size dependent rate $v_0(s)$ as defined in Eq.~\ref{eq:pden}. \textbf{(c) The attachment of proteins (shielding) alters the growth rate.} The growth rate of the mineral $v_n(s)$ depends on the size $s$ of the mineral phase as well as the number of attached FA monomers $n$. The attachment of an additional protein to a cluster shielded by $n$ proteins shields an additional surface area of size $s_{n+1}-s_{n}$. Completely-shielded particles, where the surface area $s$ is less than the shielding capacity $s_n$, do not grow.}\label{fig:fig1}
\end{figure}

For an $m$-cluster, CNT  assigns as per the \emph{capillary approximation} the free energy
\begin{equation}
\Delta G= \overbrace{\gamma f s}^{\Delta G_\gamma}+\overbrace{m\Delta\mu}^{\Delta G_\mu}= \overbrace{\gamma f s + \frac{\Delta\mu}{6\bar{v}\sqrt{\pi}}s^{3/2} }^{\textrm{spherical}} \label{eq:gibbs},
\end{equation}
where $\Delta \mu$ is the molecular free energy per monomer in the cluster 
relative to in the solution (in units $k_BT$), $\bar{v}$ is the volume
per monomer,  $\gamma>0$ is the interfacial surface energy per unit area (in units $k_BT$ per square meter), and $f$ is a geometric factor than can be adjusted to account for non-spherical growth in the pre-nucleation stage as well as size-dependent variations in the surface free energy~\cite{hu2012thermodynamics}. For constant supersaturation, $\Delta\mu<0$ so that $\Delta G\to-\infty$ as $s\to\infty$, thereby
thermodynamically favoring the existence of large clusters. Yet,  as shown
in Fig.~\ref{fig:fig1}(a),  the state of pure-monomers $(s=0)$ is also a local minimum of this free energy.
The emergence of clusters is governed by \emph{kinetic} rather than thermodynamic considerations as 
an energy barrier of  $\Delta G_{\textrm{crit}}$ corresponding to the free energy
%\begin{equation}
%\Delta G_{\textrm{crit}} = \frac{16 \pi \gamm\shat_n}{3(\rho_s\Delta\mu)^2}
%\end{equation}
of a critical cluster with size $s_*$
%\begin{equation}
%s_* =16\pi \left( \
%\frac{\gamma}{\rho_s|\Delta \mu|} \right)^2
%\end{equation}
 must be overcome.
This barrier is overcome when a cluster reaches a size $m=m_*$, $s=s_*$.
%, expressed in the surface area and volume parameterizations respectively,
%\begin{align}
% s_* &\equiv16\pi \left( \frac{\bar{v}f\gamma}{\Delta \mu}\right)^2  \label{eq:sstar} \\
% m_* &\equiv  -\frac{32\pi}{3}\left( \frac{\bar{v}\gamma f}{\Delta \mu}\right)^3.\label{eq:mstar}
%\end{align}
%
The steady-state mean-field rate at which clusters reach this size is exponential
in the magnitude of the energy gap and is known as the Zeldovich rate
\begin{equation}
j_*=\kappa \exp\left(-{\Delta G_\textrm{crit}}\right) 
\label{eq:j0}
\end{equation}
where $\kappa$ is a constant with units of concentration per time~\cite{lu2005theoretical}.
While the presence of pre-nucleation clusters in the calcium phosphate
system violates the assumptions of CNT, Habraken et al.~\cite{habraken2013ion} showed
that CNT is still applicable with the use of some minor modifications that result in
the reduction of the effective energy gap. Hence, we will assume
for the purposes of this manuscript that critically sized clusters of size $s_*$ are forming
spontaneously at some rate in the form of Eq.~\ref{eq:j0}.

  In the blood and extracellular compartments, fluid is under constant exchange. For this reason,
we will also assume that the supersaturation is constant, and hence that $j_*$ and $s_*$ are fixed, and study the growth of clusters after their nucleation.

\subsection{Growth of the mineral phase}
\label{sec:growth}

An $m$-cluster (of surface area $s(m)$) may find itself caked by a number of proteins, which effectively shield a surface area $\sprime\leq s(m)$. Our
immediate goal is to compute an effective growth rate for this particle assuming its shielding its fixed. We will assume that each successive protein
 shields a maximal surface area $\delta s_n=s_n-s_{n-1}$. In other words, if $n$ proteins are attached, then a total surface area of size $\sprime=s_n$ is shielded from further free monomer adsorption.
 
  Due to surface reactions, this particle experiences an instantaneous net flux of monomers into its structure
 \begin{equation}
J = \underbrace{k_{+}(s,\sprime)\rho_r}_{\textrm{absorption}}-\underbrace{k_{-}\times(s-\sprime)}_{\textrm{dissociation}}
\label{eq:conservation1}
\end{equation}
 where $\rho_r$ is the concentration of free monomers at the surface, $k_-$ is the dissociation rate per unit surface area, and $k_+(s,\sprime)$ is the absorption rate which is dependent on $m$ as well as the free surface area. 
 
\begin{figure}
\includegraphics[width=\columnwidth]{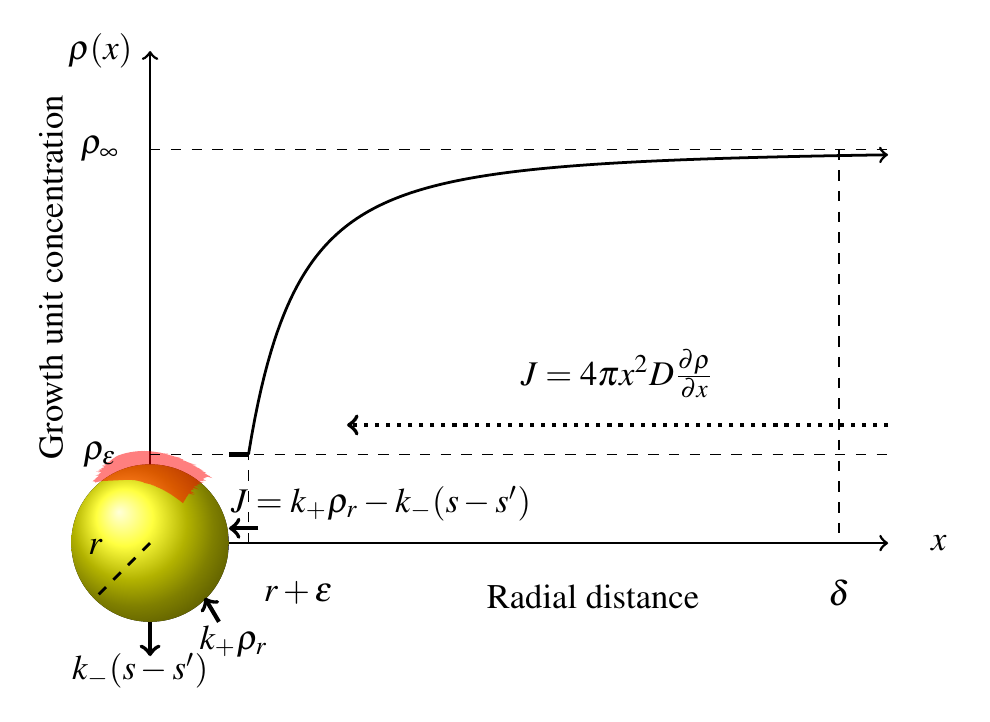}\vspace{-1em}
\caption{\textbf{Shielded diffusion-limited growth.} FA protein forms a diffusion barrier of height $\varepsilon$ and surface area $s^\prime=s_n$, where $n$ is the number of associated proteins, around the mineral cluster. The absorption
rate $k_+\rho_r$ per unit surface area of the growth units depends on the local concentration $\rho_r$ of growth units at the surface of the particle. Dissociation also occurs at rate $k_-$ per unit surface area. Neither absorption nor dissociation occur in the shielded region (red). For 
$\varepsilon$ sufficiently small, $\rho_r\approx \rho_\varepsilon$. The concentration $\rho_r$ is then determined through conservation of flux. Outside of the diffusion layer (of thickness $\delta\gg r$), the concentration of growth units approaches $\rho_\infty$.} 
\label{fig:fig2}
\end{figure}

 To begin, we will eliminate the unknown physical parameter function $k_+(s,\sprime)$ by using equilibrium considerations to relate it to the other physical parameter $k_-$. There exists a critical monomer concentration $\rho_m$ at which an $m$-cluster is at equilibrium with its surroundings so that
\begin{equation}
k_+\rho_m - k_-(s-\sprime) = 0. \label{eq:kequil0}
\end{equation}
 At the equilibrium concentration,  the free energy gap between clusters of size $m$ and $m+1$  also disappears so that,
\begin{equation}
\delta G  = \Delta \mu +(36\bar{v}^2\pi)^{1/3} \gamma( (m+1)^{2/3}-m^{2/3})  = 0,
\label{eq:kequil}
\end{equation}
where $\Delta\mu = \log(\rho_s/\rho_m)$ is the  chemical potential, and $\rho_s$ is the free monomer concentration at saturation (where the solution is in equilibrium with in infinitely-large cluster). Eq.~\ref{eq:kequil} implies that
\begin{align}
\lefteqn{\rho_m = \rho_s \exp\left[(36\bar{v}^2\pi)^{1/3} \gamma( (m+1)^{2/3}-m^{2/3})  \right] }\nonumber\\
%&= \rho_s \exp\left[ \left(\frac{32\bar{v}^2\pi}{3m}\right)^{1/3} \gamma\left( 1- \frac{1}{6m} + \mathcal{O}(m^{-2})\right)   \right] \nonumber\\
&= \rho_s \exp\left[ -\frac{\Delta\mu}{f}\left(\frac{m_*}{m}\right)^{1/3}\left(1-\frac{1}{6m} + \mathcal{O}(m^{-2})\right) \right] \nonumber\\
&= \rho_\infty \exp\left\{ \Delta\mu\left[1 -\frac{1}{f}\left(\frac{m_*}{m}\right)^{1/3} \right]\right\} \nonumber\\
&\qquad\quad \times \exp\left\{ \frac{\Delta\mu}{f}\left(\frac{m_*}{m}\right)^{1/3}\left[\frac{1}{6m} + \mathcal{O}(m^{-2})\right] \right\}.
%&= \rho_s\left[1+ \gamma\left(\frac{32\bar{v}^2\pi}{3m}\right)^{1/3} +\mathcal{O}(m^{-2/3} ) \right] \quad \textrm{as } m\to\infty.
\label{eq:rhok}
\end{align}
Substitution of $\rho_m$ from Eq.~\ref{eq:rhok} into Eq.~\ref{eq:kequil0} yields the expression for $k_+$,
\begin{equation}
k_+ =  \label{eq:kplus} \frac{k_-(s-\sprime)}{\rho_s\exp[ -\Delta\mu \sqrt{(s_*/s)}/f ]}\left(1+\mathcal{O}(m^{-4/3})\right).
\end{equation}
Eq.~\ref{eq:kplus}, substituted into Eq.~\ref{eq:conservation1}, allows us to write the flux of monomers into the mineral cluster
 \begin{align}
J \approx k_-(s-\sprime)\left[ \frac{\rho_r-\rho_s\exp\left( \frac{\beta}{\sqrt{s}} \right)}{\rho_s\exp\left(\frac{\beta}{\sqrt{s}}\right)} \right]
%&\times\left[ \frac{\rho_r-\rho_s}{\rho_s}-\frac{\rho_r(32\bar{v}^2\pi/3)^{1/3} \gamma}{\rho_sm^{1/3}}  +\mathcal{O}(m^{-2/3} )  \right], 
\label{eq:surfaceflux}
\end{align}
as a function of the physical dissociation constant $k_-$, the free monomer saturation
concentration at saturation $\rho_s$, and the free monomer concentration at the surface of the mineral $\rho_r$, and
\begin{equation}
\beta =  -\frac{\Delta\mu\sqrt{s_*}}{f}.
\end{equation}
The concentration $\rho_r$ is found through conservation of monomer mass by flux
matching as in Fig.~\ref{fig:fig2} in the quasi-steady diffusive limit, where the FA protein forms a shielding layer around the
mineral of thickness $\varepsilon$. At a distance $x$ from the center of the mineral, outside of the shielding layer ($x>r+\varepsilon$), the free monomer flux
obeys Fick's law
\begin{align}
J &= 4\pi x^2 D\partial_x\rho =  4\pi D \frac{(r+\delta)(r+\varepsilon)}{\delta -\varepsilon}(\rho_\infty -\rho_\varepsilon) \nonumber\\
&= 4\pi D(r+\varepsilon)(\rho_\infty-\rho_\varepsilon)+\mathcal{O}\left(\frac{r+\varepsilon}{\delta-\varepsilon}\right),\label{eq:bulkflux}
\end{align}
where the second equality is obtained by integration of the concentration from $x=r+\varepsilon$ to 
$x=r+\delta$, where $\delta\gg r$ is the thickness of the diffusion layer.
At the surface, the flux is given by Eq.~\ref{eq:surfaceflux}. Invoking free monomer conservation, by equating Eq.~\ref{eq:surfaceflux} with Eq.~\ref{eq:bulkflux}, while also assuming that $\varepsilon$ is small relative to the characteristic diffusion length ($\rho_r\approx\rho_\varepsilon$),
allows us to solve for the monomer concentration at the mineral surface,
\begin{equation}
\rho_r \approx \rho_s\frac{4\pi D(r+\varepsilon)\rho_\infty+k_-(s-\sprime)}{4\pi D(r+\varepsilon)\rho_s\exp\left(\frac{\beta}{\sqrt{s}}\right) +k_-(s-\sprime) }\exp\left(\frac{\beta}{\sqrt{s}}\right).
\label{eq:rhor}
\end{equation}
In Eq.~\ref{eq:rhor}, one sees that as $D\to\infty$, the concentration at $r$ goes to $\rho_\infty$, as
expected. Plugging the concentration from Eq.~\ref{eq:rhor}  into Eq.~\ref{eq:surfaceflux} yields the growth rate
\begin{align}
\dot{V} = \bar{v}J &\approx k_-(s-\sprime)\left[ \frac{4\pi\bar{v} D(r+\varepsilon)(\rho_\infty-\rho_se^{\beta/\sqrt{s}})}{4\pi D(r+\varepsilon)\rho_se^{\beta/\sqrt{s}}+k_-(s-\sprime)} \right].
\end{align}
While clusters of any size can be shielded, entirely-shielded clusters below the size $s=s_1$ are not of our concern because they are inert (recall that the first protein shields a maximal surface area of size $s_1$). Hence, we only wish to find the shielded growth rate when $s>s_1$. Making the assumption that $\varepsilon$ is small relative to $r$, for $r=\sqrt{s/4\pi} > \sqrt{ s_1 /4\pi}$, yields 
the volume growth rate in terms of $s$,
\[
\dot{V}\approx\frac{1}{\sqrt{16\pi}} \frac{\alpha s^{1/2}(s-\sprime)(1-\rho_se^{\beta/\sqrt{s}}/\rho_\infty)}{\omega s^{1/2}e^{\beta/\sqrt{s}} +(s-\sprime)},
\]
with constants 
\begin{equation}
\omega = D\sqrt{4\pi}\rho_s/k_-,\label{eq:omega}
\end{equation} 
and 
\begin{align} 
\alpha&=8\bar{v} D\pi \rho_\infty.\label{eq:alpha}
\end{align}

Note that in the limit as $s-\sprime\ll \omega s^{1/2}e^{\beta/\sqrt{s}}$, this growth rate
is surface-limited, whereas in the limit as $s-\sprime\gg \omega s^{1/2}e^{\beta/\sqrt{s}}$, growth is diffusion limited. The parameters $\omega$ and $\beta$ define the length-scale under which surface-limited effects are significant.
Typically, in nucleation problems, the surface-limited regime is ignored as it is usually only important
when the particle is small. For the purposes of our system, we take a cue from the study of~\citet{treboux2000symmetry}, which showed that aggregation of PCs is highly favorable energetically, and assume that aggregation of post-nuclear PCs is diffusion limited. However, the surface-limited effects can become significant at larger particle sizes for clusters that are nearly-completely shielded.  In this regime, we note that $e^{\beta/\sqrt{s}}\approx1$, allowing us to use the simplified growth rate
\begin{equation}
\dot{V}\approx\frac{1}{\sqrt{16\pi}} \frac{\alpha s^{1/2}(s-\sprime)}{\omega s^{1/2}  +(s-\sprime)}.
\label{eq:SVdot}
\end{equation} 
The surface area growth rate is related to the volume growth rate through the 
chain rule,
\begin{equation}
v_n(s)= \sqrt{\frac{16\pi}{s} }\dot{V}\approx\frac{\alpha(s-s_n)  }{\omega s^{1/2} +(s-s_n)}.
\end{equation}

\subsection{Shielding by the protein phase}
\label{sec:shielding}

In this section, we compute the FA shielding rate for a fixed cluster configuration. The shielding of the mineral phase by FA can be understood in a manner similar to the growth
of the mineral phase.  The overall adsorption rate of FA monomers onto the surface
 results from a balance between the diffusive supply and the surface reactions. Assuming first that mineral clusters have less mobility than FA, and denoting the diffusivity of FA by $D_{\textrm{FA}}$, one may use similar reasoning as in the previous section to find 
 the overall attachment rate of FA. As before, we may express the flux into the surface as a balance between two competing reactions through a conservation law
 \begin{align}
J_{\textrm{FA}} &= k_{\textrm{on}}\phi_r(s-\sprime) - k_{\textrm{off}}\sprime 
\label{eq:FAreactionflux}
\end{align}
where $k_{\textrm{on}}$ is the binding rate of FA to the mineral per unit free surface area per unit concentration, $k_{\textrm{off}}$ is the dissociation rate of FA, $\phi_\infty$ is the 
far-field heat bath concentration of FA, and $\phi_r$ is the concentration at the surface. By Eq.~\ref{eq:bulkflux}, we may also write the diffusive flux
\begin{equation}
J_{\textrm{FA}}\approx D_{\textrm{FA}} \sqrt{4\pi s} (\phi_\infty-\phi_r),
\label{eq:FAdiffusionflux}
\end{equation}
which, through conservation, matches the reaction flux of Eq.~\ref{eq:FAreactionflux}. Equating Eq.~\ref{eq:FAreactionflux} and Eq.~\ref{eq:FAdiffusionflux} allows us to solve for the surface concentration of FA,
\begin{equation}
\phi_r \approx \frac{ D_{\textrm{FA}} \sqrt{4\pi s} \phi_\infty + k_{\textrm{off}}\sprime }{k_{\textrm{on}}(s-\sprime)+D_{\textrm{FA}}\sqrt{4\pi s}} .
\label{eq:phi_r}
\end{equation}
Back-substituting Eq.~\ref{eq:phi_r} into Eq~\ref{eq:FAreactionflux} yields the overall rate
\begin{equation}
J_{\textrm{FA}} \approx k_{\textrm{on}}(s-\sprime)\frac{D_{\textrm{FA}}\phi_\infty\sqrt{4\pi s} + k_{\textrm{off}}\sprime}{k_{\textrm{on}}(s-\sprime)+D_{\textrm{FA}}\sqrt{4\pi s}}  - k_{\textrm{off}}\sprime.
\end{equation}
At this stage we make a few simplifying assumptions -- namely that the overall shielding process is diffusion-limited in the regime of most interest (where $0\leq \sprime < s < s_p$). First we remind the reader that our continuous formulation is an approximation of an underlying discrete system. For this reason, we note that the unshielded area $s-\sprime$ cannot become infinitesimal. This fact allows us to make the assumption that the binding reaction is always sufficiently fast such that $k_{\textrm{on}}(s-\sprime) \gg D_{\textrm{FA}}\sqrt{4\pi s} $, for $ s \leq s_p$.

 Finally, we will assume that the rate of detachments is negligible (irreversible binding of FA to mineral clusters).  Altogether, these assumptions allow us to write the simpler diffusion-limited shielding rate rule
\begin{equation}
J_{\textrm{FA}} \approx D_{\textrm{FA}}\sqrt{4\pi s}\phi_\infty.
\end{equation}

Strictly speaking, the parameters $\phi_\infty$ and $\rho_\infty$ contained in these expressions are themselves dynamical variables. Their evolution can be determined through mass conservation, as all changes are due to the balance between supply and consumption. We are interested however in the biologically-relevant situation where calcification is a local phenomenon coupled to global auto-regulatory processes that maintain supersaturation. For instance, fluid present in a knee joint is continually replenished through interstitial flow. That is to say, we set $\phi_\infty$ and $\rho_\infty$ constant and examine the conditions for the regulation of sedimentation in this regime.

\subsection{Overall continuum model}
\label{sec:continuum}

Classical work by Landau, Lifshitz~\cite{lifshitz1961kinetics},  defined an advection problem to quantitatively describe the evolution of the cluster concentrations as clusters grow due to monomer absorption.  This work has been extended throughout the years~\cite{wu1992continuum}, and recently united with nucleation~\cite{farjoun2008exhaustion,farjoun2011aggregation}, which is introduced as
 an effective boundary condition.
 We further extend this prior work by incorporating the effects of shielding. 
In this continuum approach, one may describe the evolution in size of the concentration profile
of clusters using an advection equation, where the cluster growth rate provides an
effective ``velocity'' or drift.
Overall, the dynamic concentration $c_n(s,t)$ of clusters of mineral surface area $s$ associated with $n$ FA monomers is described for all non-negative integers $n\geq 0$ for $s> s_n$ by the partial differential equations indexed by $n$,
\begin{align}
\lefteqn{\frac{\partial c_n(s,t)}{\partial t} +\frac{\partial}{\partial s}\Big[ \overbrace{\frac{\alpha(s-s_n) }{\omega s^{1/2} +(s-s_n)}}^{v_n(s)}c_n(s,t)\Big] =}\nonumber\\
&\qquad\qquad\qquad\qquad \underbrace{-\lambda s^{1/2}\left( c_n(s,t) - c_{n-1}(s,t) \right)}_{\textrm{diffusion limited}}\label{eq:pden} 
\end{align}
where $v_n(s)$ is surface growth rate dependent on the number of bound FA monomers, $\sprime = s_n$ for $n\geq 1$ and $\sprime=0$ for $n=0$, 
\begin{equation}
\lambda=\sqrt{4\pi}D_{\textrm{FA}}\phi_\infty,
\end{equation}
 and for notational convenience we set $c_{-1}(s,t)\equiv 0$.  The right-hand-side describes diffusion-limited shielding of the mineral particles by FA protein, which is assumed to have high affinity for the mineral phase. The solution domain for Eq.~\ref{eq:pden} is shown in Fig.~\ref{fig:fig1}(c). For our purposes, we will assume that the system starts at a reference time $t=t_0$ at the the initial state
\begin{equation}
c_n(s,t_0) = \begin{cases}
0 & s>s_* \\
c_0^\infty(s_*) & s= s_*,
\end{cases}\label{eq:initialcondition}
\end{equation}
where $c_0^\infty(s_*) $ is an equilibrium concentration set by the nucleation process.

Critically-sized clusters (of size $s_*$) are assumed to be created at the Zeldovich rate $j_*$ of Eq.~\ref{eq:j0}. This creation rate is balanced with consumption due to growth and shielding. As in Farjoun and Neu~\cite{farjoun2008exhaustion}, this growth flux is expressed in terms of the non-dimensional rate 
of number-growth, $\dot{V}/\bar{v}$ of Eq.~\ref{eq:SVdot}. Invoking their balance argument leads to the constraint
{
\begin{align}
j_* =&  \lim_{s\searrow s_*}\Bigg[\Bigg(D_{\textrm{FA}}\phi_\infty\sqrt{4\pi s} \nonumber\\
&\qquad\qquad+ \frac{\sqrt{4\pi} D\rho_\infty s^{3/2}}{\omega s^{1/2}+s}\Bigg)c_0(s,t) \Bigg].
\end{align}
}
Hence, the combined effects of nucleation and shielding impose an effective Dirichlet boundary condition
\begin{align}
\lefteqn{c^\infty_0(s_*) \equiv c_0(s_*,t)}\nonumber\\
&\qquad\quad= {j_*}\left[{\lambda\sqrt{s_*}+ \frac{\sqrt{4\pi} D \rho_\infty s_*^{3/2}}{\omega s_*^{1/2} +s_*}}\right]^{-1}.
\label{eq:bc}
\end{align}

\section{Results}

In this section we construct a solution to the system of partial differential equations defined
in Eq.~\ref{eq:pden} and the boundary conditions defined in Eqs.~\ref{eq:initialcondition}--\ref{eq:bc}. We proceed first by non-dimensionalization of the problem formulation. Then, using
the method of characteristics, we derive a sequential relationship between the solutions of the system. Approximating the solution of the system of equations, we analyze the steady-state behavior of the system in the limit where inhibition is sufficiently strong, from which we compute the overall rate of protein consumption and propose a criterion for effective inhibition of calcification. Finally, we parameterize our equations using experimentally measured values found in the literature.

\subsection{Nondimensionalization}
We seek a convenient non-dimensionalization of our serial system of PDEs describing the
shielded growth problem. We begin by normalizing the surface area $s$, which ranges between
the critical nucleation surface area $s_*$ and another critical surface area $s_p$ which represents the
surface area at sedimentation. Using these constants, we write the non-dimensionalized
size variable
\begin{equation}
\hat{s} = \frac{s-s_*}{s_p-s_*}
\end{equation}
 critical cluster size $s_*$,
\begin{equation}
\shat_{*} = \frac{s_*}{s_p-s_*},
\end{equation}
shielded surface area,
\begin{equation}
\hat{s}^\prime= \frac{s^\prime}{s_p-s_*},
\end{equation}
shielding levels $\sprime=s_n$,
\begin{equation}
\shat_n= \frac{s_n}{s_p-s_*},
\end{equation}
and surface-limiting parameter $\omega$
\begin{equation}
\hat{\omega} = \frac{\omega}{\sqrt{s_p-s_*}}.
\end{equation}
Rescaling time
\begin{equation}
\hat{t} = \frac{\alpha (t-t_0)}{s_p-s_*}
\end{equation}
results in the series of non-dimensional advection equations of asymptotically unit speed,
\begin{widetext}
\begin{align}
\frac{\partial \hat{c}_n}{\partial \hat{t}} + \frac{\partial}{\partial \hat{s}}\left[\frac{(\hat{s}+\shat_{*}-\hat{s}_n)\hat{c}_n(\hat{s},\hat{t})}{\hat{\omega}(\hat{s}+\shat_{*})^{1/2}+(\hat{s}+\shat_{*}-\hat{s}_n)} \right] = -\hat{\lambda}\sqrt{\hat{s}+\shat_{*}}\left(  \hat{c}_n(\hat{s},\hat{t}) - \hat{c}_{n-1}(\hat{s},\hat{t})  \right)\label{eq:ndpde}
\end{align}
\end{widetext}
with non-dimensional shielding constant
\begin{align}
\hat{\lambda} &= \frac{\lambda(s_p-s_*)^{3/2}}{\alpha}
\end{align}
where the concentrations have been scaled by the nucleation boundary condition 
\begin{equation}
\hat{c}_n(\hat{s},\hat{t}) =\frac{  c_n(s(\hat{s}),t(\hat{t})) }{c^\infty_0(s_*)}, 
\end{equation}
so that the concentration of critical clusters is fixed
\begin{equation}
\hat{c}_0(0,\that) =1.
\end{equation}

\subsection{Characteristics of the PDE system}

\begin{figure}[!t]
\includegraphics[width=0.9\columnwidth]{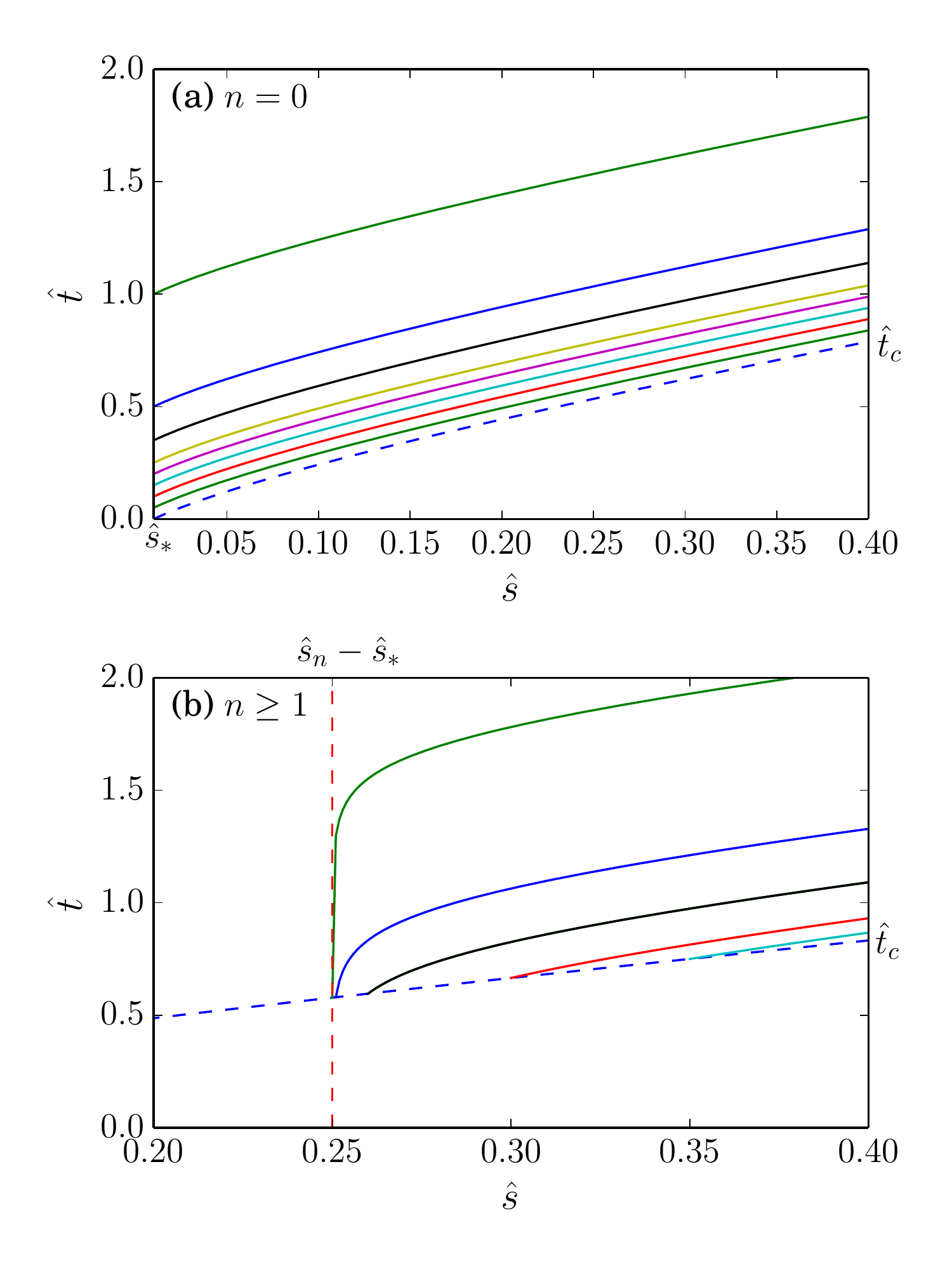}
\caption{\textbf{Characteristics for advection problem.} Some sample $\shat-\that$ characteristics for the nondimensionalized PDE problem with $\shat_n-\shat_*=0.25$. For (a) $n=0$, and characteristics emerge from $\shat=\shat_*$. For (b) $n\geq 1$,  the characteristics emerge from the curve $\{(\shat,\that_c(\shat))\}$, where $\that_c$ is given in Eq.~\ref{eq:tc}. }
\label{fig:characteristics}
\end{figure}

The non-dimensionalized partial differential equations of Eq.~\ref{eq:ndpde} can be solved by invoking the method of characteristics sequentially for each PDE. 
%In our solution we will assume that $\beta/\sqrt{s_*}\ll 1$, so that we may eliminate the exponential terms.  We do so because $\rho_s$ is a multiplicative pre-factor that appears anyplace where $\exp(-\beta/\sqrt{s_*})$ appears, and $\rho_s/\rho_\infty$ would be exponentially small in any situation where $\exp(-\beta/\sqrt{s_*})$ is small. Conversely, it is possible in some systems for $\exp(-\beta/\sqrt{s_*})\approx 1$ to hold whereas $\omega$ is not negligible.  Hence, we retain $\omega$ in our formulation to accommodate descriptions of these systems. 
%
The solutions to the PDEs
contain the characteristic curves described by the equations
\begin{align}
\frac{\dd \shat}{\dd \that} &= \frac{(\hat{s}+\shat_{*}-\hat{s}^\prime) }{\hat{\omega}(\hat{s}+\shat_{*})^{1/2}+(\hat{s}+\shat_{*}-\hat{s}^\prime)}.\label{eq:dsdt}
\end{align}
The $\hat{s}-\hat{t}$ characteristics, as shown in Fig.~\ref{fig:characteristics}, originate from
points $(\that_0,\shat(\that_0))$. They follow the relationship
%
%In the regime that $\hat{\beta}/\sqrt{s}\ll1$, this relationship becomes
\begin{align}
\hat{t}-\hat{t}_0 &= \hat{s}-\hat{s}(\hat{t}_0) +2\hat{\omega}\left[ \sqrt{\hat{s}+\shat_{*}}-\sqrt{\hat{s}(\hat{t}_0)+\shat_{*}} \right] \nonumber\\
&+\hat{\omega}\shat^\prime\log\left(\frac{\sqrt{\hat{s}+\shat_{*}}-\sqrt{\shat^\prime}}{\sqrt{\hat{s}+\shat_{*}}+\sqrt{\shat^\prime}} \frac{\sqrt{\hat{s}(\hat{t}_0)+\shat_{*}}+\sqrt{\shat^\prime}}{\sqrt{\hat{s}(\hat{t}_0)+\shat_{*}}-\sqrt{\shat^\prime}} \right).\label{eq:stcurves}
\end{align}
Particularly, for unshielded clusters (where $\shat^\prime=0$), the last line of Eq.~\ref{eq:stcurves} is zero.

Along these curves, the concentration varies as
\begin{align}
\frac{\dd\hat{c}_n }{\dd\hat{t}} &= -\frac{\hat{\omega}}{2\sqrt{\hat{s}+\shat_{*}}}\frac{\hat{s}+\shat_{*}+\hat{s}^\prime}{\left[ \hat{\omega}\sqrt{\hat{s}+\shat_{*}} + \hat{s}+\shat_{*}-\hat{s}^\prime  \right]^2}\hat{c}_n(\hat{t}) \nonumber\\
&\qquad-\hat{\lambda}\sqrt{\hat{s}+\shat_{*}}\left(  \hat{c}_n(\hat{t}) - \hat{c}_{n-1}(\shat,\hat{t})  \right).\label{eq:cn}
\end{align} 
For the purpose of solving these equations, it is advantageous to invoke the change-of-variables 
\begin{align}
u&\equiv\sqrt{\shat+\shat_{*}} \\
\hat{s}^\prime&\equiv \shat_n \\ 
\dd{s}&=2u\dd{u},
\end{align}
 to reparameterize the curves as
\begin{align}
\frac{\dd \that}{\dd u} &= 2u\left[\frac{\hat{\omega} u}{u^2-\shat_n}+1\right]\geq 0\qquad \forall u > \sqrt{\shat_n},
\end{align}
from where it is evident that the relationship between $\that$ and $u$ is bijective. Hence, we may use $u$
as a proxy for $\that$, finding that the concentration profiles along these curves vary with $u$ as 
\begin{align}
\lefteqn{\frac{\dd \hat{c}_n}{\dd u} = } \nonumber\\
&\qquad-\left[\frac{\hat{\omega}}{u^2-\shat_n}\frac{u^2+\shat_n}{\hat{\omega} u + u^2-\shat_n} + 2\hat{\lambda} u^2\frac{u^2+\hat{\omega} u -\shat_n}{u^2-\shat_n} \right]\hat{c}_n \nonumber\\
&\qquad\qquad \qquad\qquad+2\hat{\lambda} u^2\frac{u^2+\hat{\omega} u -\shat_n}{u^2-\shat_n} \hat{c}_{n-1}(\shat(u),\that(u)).\label{eq:dcdu}
\end{align}
With the aid of an integrating factor, Eq.~\ref{eq:dcdu} can be written in the exact differential form
\begin{align}
\lefteqn{\dd\left\{ \frac{ (u^2-\shat_n)^{\hat{\lambda}\hat{\omega} \shat_n+1}}{u^2+\hat{\omega} u -\shat_n}\exp\left[\hat{\lambda} u^2\left(\hat{\omega}+\frac{2}{3}u \right) \right] \hat{c}_n\right\} }\nonumber\\
&= 2\hat{\lambda} u^2(u^2-\shat_n)^{\hat{\lambda}\hat{\omega} \shat_n}\exp\left[\hat{\lambda} u^2\left(\hat{\omega}+\frac{2}{3}u \right) \right] \hat{c}_{n-1}(u,\that(u)).\label{eq:exactdiff}
\end{align}
The equation for the unshielded particle concentration $\hat{c}_0$ corresponds to the homogeneous problem $(\shat_0\equiv 0;\ \hat{c}_{-1}\equiv0)$. Respecting the boundary condition invoked by nucleation, as
well as the initial Cauchy data, yields the solution
\begin{align}
\lefteqn{\hat{c}_0(\shat,\that)  = H\left(\that-\that_c(\shat)\right)\frac{\hat{\omega}+\sqrt{\shat+\shat_{*}}}{\sqrt{\shat+\shat_{*}}}\frac{\sqrt{\shat_{*}}}{\hat{\omega}+\sqrt{\shat_{*}}} }\nonumber\\
&\qquad\qquad\qquad\qquad\times\frac{\exp\left[\hat{\lambda} \shat_{*}\left(\hat{\omega}+\frac{2}{3}\sqrt{\shat_{*}} \right) \right]}{\exp\left[\hat{\lambda} (\shat+\shat_{*})\left(\hat{\omega}+\frac{2}{3}\sqrt{\shat+\shat_{*}} \right) \right]} ,\label{eq:c0}
\end{align}
where $H$ is the unit step function and 
\begin{equation}
\that_c(\shat) = \shat + 2\hat{\omega}\left(\sqrt{\shat+\shat_{*}}-\sqrt{\shat_{*}}  \right)\label{eq:tc}
\end{equation}
is analogous to a ``first-passage-time'' for the formation of size-$\shat$ clusters.

To solve for the subsequent concentrations, we take advantage of the Cauchy data by 
initializing all characteristic curves along the curve $(\that_c(\shat),\shat)$ given by Eq.~\ref{eq:tc}, thereby
setting $\hat{c}_n=0$ at the left endpoint. Hence, each point $(\shat,\that)$ such that $\shat\geq \hat{s}^\prime=\shat_n$,
and $\that \geq \that_c(\shat)$ lies uniquely on a single curve originating from
\begin{align}
\lefteqn{\sqrt{\shat(\that_0)+\shat_{*}}= }\nonumber\\
&\quad\sqrt{\shat^\prime}\frac{\frac{\sqrt{\shat+\shat_{*}} +\sqrt{\shat^\prime}}{\sqrt{\shat+\shat_{*}} -\sqrt{\shat^\prime}}\exp\left[\frac{\that-\shat-2\hat{\omega}(\sqrt{\shat+\shat_{*}}-\sqrt{\shat_{*}})}{\hat{\omega}\shat^\prime} \right]  +1 }{ \frac{\sqrt{\shat+\shat_{*}} +\sqrt{\shat^\prime}}{\sqrt{\shat+\shat_{*}} -\sqrt{\shat^\prime}}\exp\left[\frac{\that-\shat-2\hat{\omega}(\sqrt{\shat+\shat_{*}}-\sqrt{\shat_{*}})}{\hat{\omega}\shat^\prime} \right]-1} .\label{eq:st0}
\end{align}

As growth of the mineral phase occurs more quickly in the
unshielded clusters than in the shielded clusters, the $\shat-\that$ characteristics propagate
quickest in the unshielded clusters. Hence, the hierarchy  
$\textrm{supp}(\hat{c}_n)\subseteq\textrm{supp}(\hat{c}_{n-1}) \subseteq\cdots\subseteq\textrm{supp}(\hat{c}_0)$ holds. In fact, the supports of all functions $\hat{c}_n$ are equal, as necessitated by the coupling defined by the right-hand-side of Eq.~\ref{eq:ndpde}. The creation of size-$\shat$ clusters of shielding $n$ is driven more by the shielding of ``$n-1$ clusters'' rather than the growth of ``$n$ clusters.''
We use this fact, along with the presence of an exponential term within the exact differential of
the right hand side of Eq.~\ref{eq:exactdiff} to formulate the ansatz
\begin{widetext}
\begin{align}
\hat{c}_n(\shat,\that) = H\left(\that-\that_c(\shat)\right)g_n(\shat,\that)\exp\left[-\hat{\lambda} (\shat+\shat_{*})\left(\hat{\omega}+\frac{2}{3}\sqrt{\shat+\shat_{*}} \right) \right] \exp\left[\hat{\lambda} \shat_{*}\left(\hat{\omega}+\frac{2}{3}\sqrt{\shat_{*}} \right) \right],\label{eq:chatn}
\end{align}
\end{widetext}
where
\begin{align}
\lefteqn{g_n(u) = 2\hat{\lambda}\frac{u^2+\hat{\omega} u-\shat_n}{(u^2-\shat_n)^{\hat{\lambda} \hat{\omega} \shat_n+1}} } \nonumber\\
&\qquad\qquad\times\int_{\sqrt{\shat(\that_0)+\shat_{*}}}^u q^2(q^2-\shat_n)^{\hat{\lambda}\hat{\omega} \shat_n}g_{n-1}(q)\dd{q}.
\label{eq:fnu}
\end{align}

According to Eq.~\ref{eq:c0},
\begin{equation}
g_0(u) = \frac{\hat{\omega}+u}{u}\frac{\sqrt{\shat_{*}}}{\hat{\omega}+\sqrt{\shat_{*}}} .
\label{eq:f0u}
\end{equation}
For solving $g_n(\shat,\that)$, the lower bound for the integral in Eq.~\ref{eq:fnu} is taken from Eq.~\ref{eq:st0}. For solving the next equation $g_{n+1}(\shat,\that)$, all instances of $\shat,\that$ reparameterized by the variable $u$ using Eq.~\ref{eq:st0}. 

The iterated integrals of Eq.~\ref{eq:fnu} can be solved numerically through standard quadrature methods. Here we find some properties of the solutions to these equations before exploring their steady-state behavior, which is of the most interest to us.

First, there is the question of whether these equations are well-posed. For $u$ near $\sqrt{\shat_n}$, one can invoke L'Hopital's rule on Eq.~\ref{eq:fnu} to find that 

\[
\lim_{u\searrow \sqrt{\shat_n}} g_n(u) = \frac{\hat{\lambda}\hat{\omega}\sqrt{\shat_n}}{\hat{\lambda}\hat{\omega}\shat_n +1} = \mathcal{O}(1).
\]

So, the solutions are bounded on the left. Now, we seek to find pointwise bounds for the solution away from the left boundary (for $u>\sqrt{\shat_n}$). We note that the lower bound of the integral term in Eq.~\ref{eq:fnu}, given by Eq.~\ref{eq:st0}, approaches $\sqrt{\shat_n}$ as $t\to\infty$. Since the integrand is non-negative, the solution is bounded from above by the steady state solution
\begin{align}
g_n(u) &= 2\hat{\lambda}\frac{u^2+\hat{\omega} u-\shat_n}{(u^2-\shat_n)^{\hat{\lambda} \hat{\omega} \shat_n+1}} \nonumber\\
&\qquad\times\int_{\sqrt{\shat(\that_0)+\shat_{*}}}^u q^2(q^2-\shat_n)^{\hat{\lambda}\hat{\omega} \shat_n}g_{n-1}(q)\dd{q} \nonumber\\
&\leq 2\hat{\lambda}\frac{u^2+\hat{\omega} u-\shat_n}{(u^2-\shat_n)^{\hat{\lambda} \hat{\omega} \shat_n+1}} \nonumber\\
&\qquad\times\int_{\sqrt{\shat_n}}^u q^2(q^2-\shat_n)^{\hat{\lambda}\hat{\omega} \shat_n}g_{n-1}(q)\dd{q} \nonumber \\
&\equiv g_n^\infty(u). \label{eq:fnu_upper_bound}
\end{align}

By repeated applications of the Cauchy-Schwarz inequality, one sees that the integral term in Eq.~\ref{eq:fnu_upper_bound} satisfies the inequalities
\begin{align*}
\lefteqn{\int_{\sqrt{\shat_n}}^u q^2(q^2-\shat_n)^{\hat{\lambda}\hat{\omega}\shat_n} g_{n-1}(q)\dd q }\nonumber \\
&\leq\left[ \int_{\sqrt{\shat_n}}^u q^4 (q^2-\shat_n)^{2\hat{\lambda}\hat{\omega}\shat_n}\dd q \right]^{1/2} || g^\infty_{n-1} ||_{L^2(\sqrt{\shat_n},u)} \nonumber\\
&\leq\left[ \frac{(u^2-s_n)^{4\hat{\lambda}\hat{\omega}\shat_n+1} }{4\hat{\lambda}\hat{\omega}\shat_n+1} \right]^{1/4}\left[\frac{u^8-s_n^4}{8}  \right]^{1/4}   || g_{n-1}||_{L^2(\sqrt{\shat_n},u)}.
\end{align*}
This computation gives us the pointwise bound on $g_n^\infty$,
\begin{align*}
g_n^\infty(u)&\leq \frac{2\hat\lambda}{(4\hat\lambda\hat\omega\shat_n+1)^{1/4}} \frac{u^2+\hat{\omega} u-\shat_n}{ (u^2-\shat_n)^{3/4}}\nonumber\\
&\qquad\times\left[\frac{u^8-\shat_n^4}{8}  \right]^{1/4}   || g_{n-1}||_{L^2(\sqrt{\shat_n},u)}.
\end{align*}
Using the fact that $g^\infty_0(u)= \mathcal{O}(1)$, it is easy to see by induction that $g_n(u)$ is bounded and smooth (behaving locally like a polynomial with order controlled by $n$) for $u>\sqrt{\shat_n}$, where it is of-note that $g_n^\infty$ is bounded also in the vicinity of $\sqrt{\shat_n}$. Hence, by Eq.~\ref{eq:chatn}, each solution of $c_n(s,t)$ is Schwartz-class, for all $t\geq 0$.

\subsection{Steady-state behavior}

Our interest is in long-term behavior of the system. Observe that the solutions of Eq.~\ref{eq:chatn} contain an exponential multiplicative factor that represents regulatory shielding. This shielding is \emph{strong} provided that the term in the exponential is large, which is the case when the inhibition ratio obeys
\begin{equation}
\hat{\lambda}(1+\shat_*)^{3/2}\gg 1.\label{eq:inhibitionratio}
\end{equation}
In physical units, this criterion can be expressed succinctly in terms of the concentration of FA protein needed,
\begin{equation}
\phi_\infty \gg \frac{D}{D_{FA}} \frac{\rho_\infty-\rho_s}{m_p},
\label{eq:strongcondition}
\end{equation}
where $m_p$ is critical number of Posner clusters in a pure-mineral cluster at sedimentation.  Note that if this condition were not to hold then a significant number of clusters of sedimentation size would form. Sedimentation would then occur until exhaustion of supersaturated species.
 In our subsequent analysis, we will assume that this condition holds. 
 
Since the overall solution is tapered by the exponential term which goes as $s^{3/2}$, or as the volume, we are most interested in the behavior of $g_n^\infty(u)$ in the vicinity of $\sqrt{s_n}$.  In this limit, we use the binomial theorem to approximate the integrals of the general form, for $a,b\in\mathbb{R}$,
{\small
\begin{align}
\lefteqn{\int_{\sqrt{\shat_n}}^u q^2(q^2-\shat_n)^{\hat{\lambda}\hat{\omega}\shat_n} q^a(q^2-\shat_{n})^b\dd{q} } \nonumber\\
&=\frac{\sqrt{\shat_n^{a+1}}}{2} \int_{0}^{u^2-\shat_n} x^{\hat{\lambda}\hat{\omega}\shat_n+b}\left(1+\frac{x}{s_n} \right)^{(a+1)/2}\dd x \nonumber\\
&=\frac{\sqrt{\shat_n^{a+1}}}{2} \left[\frac{(u^2-\shat_n)^{\hat{\lambda}\hat{\omega}\shat_n+b+1}}{\hat{\lambda}\hat{\omega}\shat_n+b+1} +\mathcal{O}\left((u^2-\shat_n)^{\hat{\lambda}\hat{\omega}\shat_n+b+2} \right)  \right]. \label{eq:genintegral}
\end{align}}

Eq.~\ref{eq:genintegral} allows us to evaluate $g^\infty_1$ to the leading order
\begin{align}
g^\infty_1(u) &\approx \frac{\sqrt{\shat_*}}{\hat\omega+\sqrt{\shat_*}} \frac{\hat\lambda(\sqrt{\shat_1}+\hat\omega)}{\hat{\lambda}\hat{\omega}\shat_1+1}\left[  (u^2-\shat_1)+\hat\omega u \right]. \nonumber\\
%&= \frac{\hat\lambda\sqrt{\shat_1+\hat\omega}}{\hat{\lambda}\hat{\omega}\shat_1+1}\left[  (u^2-\shat_2) + \shat_2-\shat_1+\hat\omega u \right]
\end{align}
Through an inductive argument, one finds that for $n\geq 1$,
\begin{align}
\lefteqn{g_n^\infty(u) \approx (u^2-\shat_n+\hat\omega u)\frac{\sqrt{\shat_*}}{\hat\omega+\sqrt{\shat_*}} } \nonumber\\
&\qquad\qquad\times\frac{1}{\shat_1}\prod_{j=1}^n\frac{\hat\lambda( (\shat_j-\shat_{j-1})\sqrt{\shat_j} +\shat_j\hat\omega)}{\hat\lambda\hat\omega\shat_j+1}.
\end{align}
We retain both $u^2-s_n$ and $\hat\omega u$ in this expression because it is unclear which term is large. As $n$ increases however, the $\hat\omega u$ term will begin to dominate. This fact implies that surface-limited effects arise for large clusters, contrary to the situation for most other nucleation problems.

\begin{figure}
\includegraphics[width=0.75\columnwidth]{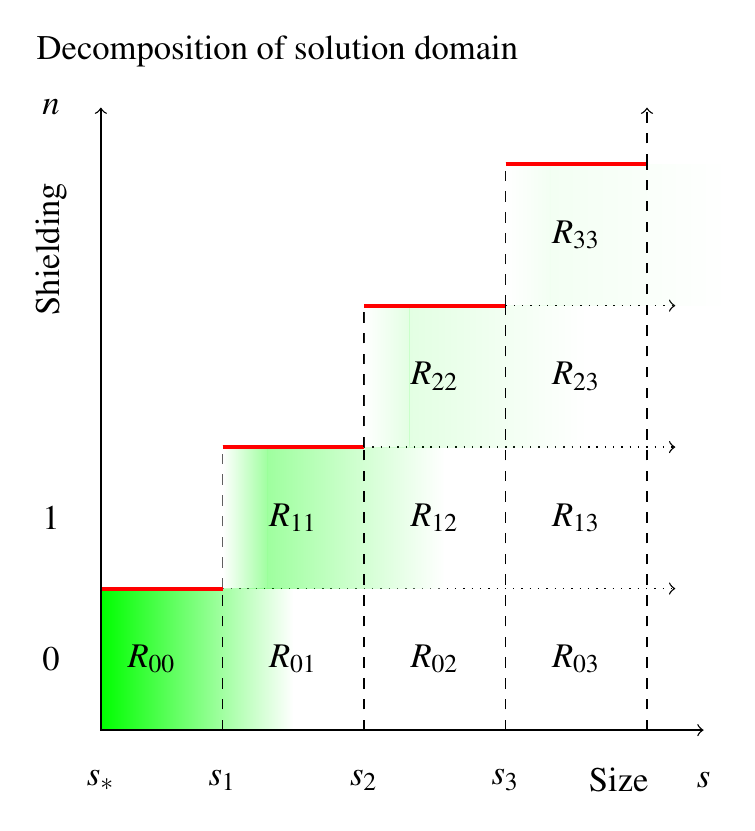}
\caption{\textbf{Steady-state cluster concentration $c^\infty_n(s)$ and domain decomposition for computing the protein consumption rate.} \textit{Green shading (darker is more concentrated) corresponds to increased steady-state concentration of mineral clusters of the particular size (given by horizontal axis) with the given number of attached FA proteins (given by the vertical axis).} In the asymptotic case of strong shielding (Eq.~\ref{eq:strongcondition}), an exponential decay of concentration is seen according to size. From this solution, an overall consumption rate of FA protein can be computed by summing over the attachment rates $R_{nj}$, where $R_{nj}$ refers to the rate of protein consumption by clusters of $n$ bound proteins of size $s_j$ to $s_{j+1}$. }
\label{fig:solution}
\end{figure}

\subsection{Rate of FA consumption}

Since we know the rate of protein association as a function of $n$ (the number of bound proteins) and $s$ (the surface area of the mineral phase), we can compute the total rate of FA consumption. Denote $R_{jk}$ the cumulative rate of mineral consumption in shielding particles of $j$ FA monomers with size $s\in (s_k,s_{k+1})$. We derive this rate first  for the shielding of unshielded clusters of size at most $s_1$, $R_{00}$.
Returning back to an integer parameterization of the size, it is clear that the total rate of consumption of FA for these clusters follows
\begin{align}
R_{00}=\sum_{m=m(s_*)}^{m(s_1)} \lambda\sqrt{s(m)} c_0^\infty(s(m)).\label{eq:R001}
\end{align}
The sum in Eq.~\ref{eq:R001} can  be approximated by a left-Riemann integral so that
\begin{align}
\lefteqn{\sum_{m=m(s_*)}^{m(s_1)} \lambda\sqrt{s(m)} c_0^\infty(s(m)) \approx}\nonumber\\
&\qquad\qquad  \int_{m(s_*)}^{m(s_1)} \lambda\sqrt{s(m)} c_0^\infty(s(m))\dd{m}.
\end{align}
After transformation from $m$ back to $s$, one finds that 
\begin{align}
R_{00} \approx \frac{\lambda}{\sqrt{16\bar{v}^2\pi}} \int_{s_*}^{s_1}  s c^{\infty}_0(s)\dd s.
\end{align}
Generalizing this result, it is easy to see that
\begin{equation}
R_{nj} \approx \frac{\lambda}{\sqrt{16\bar{v}^2\pi}} \int_{s_j}^{s_{j+1}}  s c^{\infty}_n(s)\dd s.
\label{eq:Rnj}
\end{equation}
The total consumption rate of FA protein follows
\begin{equation}
R\approx\overbrace{\lambda\sqrt{s_*}c_0^\infty(s_*)}^{\textrm{at nucleation}}+ \sum_{n=0}^\infty \sum_{j=n}^\infty R_{nj}  
\end{equation}
where the first term represents FA consumed in the instantaneous shielding of critical clusters. Substituting Eq.~\ref{eq:chatn} yields
 \begin{align}
\lefteqn{R\approx\lambda\sqrt{s_*}c_0^\infty(s_*)\Bigg\{1+\exp\left[ \frac{\lambda s_*}{\alpha}\left(\omega+\frac{2}{3}\sqrt{s_*}\right)  \right] } \nonumber\\
&\quad\times\sum_{n=0}^\infty \int_{s_n}^\infty \frac{ sg_n^\infty\left(u(s)\right)}{\sqrt{16\bar{v}^2s_*\pi}} 
\exp\left[-\frac{\lambda s}{\alpha}\left(\omega +\frac{2}{3} \sqrt{s} \right)  \right] \dd s  \Bigg\},
\label{eq:R}
\end{align}
where $u(s)=\sqrt{s/(s_p-s_*)}$. We approximate the integrals in Eq.~\ref{eq:R} using the change of variables
\begin{align*}
\xi &= \frac{\lambda s}{\alpha}\left(\omega+\frac{2}{3}\sqrt{s}\right),
\end{align*}
which we invert to find that
\begin{align}
s &= \left(\frac{3\alpha}{2\lambda}\xi \right)^{2/3}- \omega \left(\frac{3\alpha}{2\lambda}\xi \right)^{1/3} + \mathcal{O}
(\omega^2) \nonumber\\
\dd\xi &= \frac{\lambda}{\alpha}\left(\omega+\sqrt{s}\right)\dd s \nonumber\\
&=\frac{\lambda}{\alpha}\left[ \left(\frac{3\alpha}{2\lambda}\xi \right)^{1/3} + \frac{\omega}{2} + \mathcal{O}(\omega^2) \right] \dd s.
\end{align}
For $n\geq1$, we evaluate the integrals in Eq.~\ref{eq:R},
\begin{align}
\lefteqn{\int_{s_n}^\infty\frac{ sg_n^\infty\left(u(s)\right)}{\sqrt{16\bar{v}^2s_*\pi}}
\exp\left[-\frac{\lambda s}{\alpha}\left(\omega +\frac{2}{3} \sqrt{s} \right)  \right] \dd s  } \nonumber \\
&=\frac{1}{\sqrt{16\bar{v}^2s_*\pi}} \frac{\sqrt{s_*}}{\omega+\sqrt{s_*}}\frac{1}{s_1}\prod_{j=1}^n\frac{\lambda( (s_j-s_{j-1})\sqrt{s_j} + s_j\omega  )}{\lambda\omega s_j +\alpha}  \nonumber\\
&\quad \times \int^\infty_{ \frac{\lambda s_n}{\alpha}\left(\omega+\frac{2}{3}\sqrt{s_n}\right)} \frac{s(s-s_n + \omega\sqrt{s})}{\frac{\lambda}{\alpha}\left( \left(\frac{3\alpha}{2\lambda}\xi \right)^{1/3} + \frac{\omega}{2} + \mathcal{O}(\omega^2) \right) } e^{-\xi} \dd \xi. \label{eq:integral0}
\end{align}
We approximate the integral in Eq.~\ref{eq:integral0} by a sum of incomplete Gamma functions $\Gamma(a,x)$,
\begin{align}
\lefteqn{\int^\infty_{ \frac{\lambda s_n}{\alpha}\left(\omega+\frac{2}{3}\sqrt{s_n}\right)} \frac{s(s-s_n + \omega\sqrt{s})}{\frac{\lambda}{\alpha}\left( \left(\frac{3\alpha}{2\lambda}\xi \right)^{1/3} + \frac{\omega}{2} + \mathcal{O}(\omega^2) \right) } e^{-\xi} \dd \xi } \nonumber\\
&\quad=\frac{\alpha}{\lambda}\int^\infty_{ \frac{\lambda s_n}{\alpha}\left(\omega+\frac{2}{3}\sqrt{s_n}\right)}\Bigg\{ \left(\frac{3\alpha}{2\lambda}\xi \right) + s_n\left(\frac{3\alpha}{2\lambda}\xi \right)^{1/3} \nonumber\\
&\quad\qquad-{\omega}\left[\frac{5}{2}\left(\frac{3\alpha}{2\lambda}\xi \right)^{2/3}-\frac{s_n}{2}  \right] +\mathcal{O}(\omega^2)\Bigg\}e^{-\xi} \dd \xi \nonumber\\
&\quad\sim  \frac{3\alpha^2}{2\lambda^2} \Gamma\left( 2,  \frac{\lambda s_n}{\alpha}\left(\omega+\frac{2}{3}\sqrt{s_n}\right) \right) \nonumber\\
&\qquad\qquad+ s_n\left(\frac{3\alpha^4}{2\lambda^4}\right)^{1/3} \Gamma\left(\frac{4}{3},  \frac{\lambda s_n}{\alpha}\left(\omega+\frac{2}{3}\sqrt{s_n}\right)   \right)\nonumber\\
&\qquad\qquad-\frac{5s_n \alpha\omega}{ 2\lambda}\left(\frac{3\alpha}{2\lambda}\right)^{2/3} \Gamma\left(\frac{5}{3} , \frac{\lambda s_n}{\alpha}\left(\omega+\frac{2}{3}\sqrt{s_n}\right)  \right)\nonumber\\
&\qquad\qquad +\frac{\alpha s_n\omega}{2\lambda} \Gamma\left( 1,  \frac{\lambda s_n}{\alpha}\left(\omega+\frac{2}{3}\sqrt{s_n}\right)  \right),
\end{align}
where the remainder term is exponentially small by rationale of Watson's Lemma. For $n=0$, we have
\begin{align}
\lefteqn{\int_{s_*}^\infty \frac{ sg_0^\infty\left(u(s)\right)}{\sqrt{16\bar{v}^2s_*\pi}} 
\exp\left[-\frac{\lambda s}{\alpha}\left(\omega +\frac{2}{3} \sqrt{s} \right)  \right] \dd s } \nonumber\\
&= \frac{ 1}{\sqrt{16\bar{v}^2s_*\pi}}\frac{\sqrt{s_*}}{\omega+\sqrt{s_*}} \nonumber\\
&\quad\times\int_{s_*}^\infty \left(s+{\omega}{\sqrt{s}}\right) \exp\left[-\frac{\lambda s}{\alpha}\left(\omega +\frac{2}{3} \sqrt{s} \right)  \right] \dd s  \\
& =  \frac{ 1}{\sqrt{16\bar{v}^2s_*\pi}}\frac{\sqrt{s_*}}{\omega+\sqrt{s_*}} \nonumber\\
&\quad\times \frac{\alpha}{\lambda} \int_{ \frac{\lambda s_*}{\alpha}\left(\omega+\frac{2\sqrt{s_*}}{3}\right)}^\infty\left[ \left(\frac{3\alpha}{2\lambda}\xi \right)^{1/3} -\frac{\omega}{2}+\mathcal{O}(\omega^2) \right] e^{-\xi}\dd\xi\nonumber \\
&\sim  \frac{ \alpha}{\lambda\sqrt{16\bar{v}^2s_*\pi}}\frac{\sqrt{s_*}}{\omega+\sqrt{s_*}} \Bigg\{\left(\frac{3\alpha}{2\lambda} \right)^{1/3}\Gamma\left( \frac{4}{3}, \frac{\lambda s_*}{\alpha}\left(\omega+\frac{2\sqrt{s_*}}{3} \right)  \right) \nonumber\\
&\qquad\qquad\qquad-\frac{\omega}{2} \Gamma\left(1, \frac{\lambda s_*}{\alpha}\left(\omega+\frac{2\sqrt{s_*}}{3} \right)  \right)  \Bigg\}.
\end{align}

In the regime where $\lambda/\alpha\to\infty$, no clusters of size greater than $s_*$ form, and the rate of FA consumption matches the nucleation rate for ACP. Conversely, if clusters of size greater than $s_1$ form, the FA consumption rate is strictly greater than the ACP nucleation rate.

\subsection{Parameterization}

As we have mentioned, normal physiological calcium concentrations  exceed  supersaturation relative to the most thermodynamically stable phase of calcium phosphate. In fact, as  we shall see, they also exceed supersaturation relative to ACP.

Normal serum free ionic \ce{Ca^2+} concentration varies between $1.2$mM and $1.3$mM~\cite{carrol2003practical}, and normal total serum  phosphate concentration varies between $1.12$mM and $1.45$mM. At $\textrm{pH}=7.4$, one finds using the Henderson-Hasselbalch equation that the concentration of free \ce{PO4^3-} lies between $3.7$ and $4.9$ nM.
Various studies have explored the solubility of ACP relative to concentrations of its constituent ions (\ce{Ca^2+}, \ce{PO4^3-}). By empirical formula, ACP has the negative-log$_{10}$-solubility $\textrm{p}K_s  = 3\textrm{p}\ce{Ca} + 2\textrm{p}\ce{PO4}\approx 26$ at 310K and pH$=7.4$~\cite{christoffersen1990apparent}. Using this calculation, one may compute the supersaturation relative to ACP,
\begin{equation}
S^{(0)}=\left(\frac{\ce{[Ca^2+]^3 [PO4^3-]^2}}{10^{-26}}\right)^{1/5}.
\end{equation}
Hence, the supersaturation ratio is approximately $S^{(0)}\approx1.3$ in normal conditions.

Estimates for the molecular weight of FA range from $51-67$kDa, while the usual serum concentrations of FA range from $0.5-1.0$g/L. Thus, FA ranges in concentration between $7\mu$M$-19\mu$M in normal situations.  We assume that $m_p=\mathcal{O}(10^2)$, where it 
notable that a single FA protein shields approximately $10^2$ Posner clusters~\cite{heiss2010fetuin}. For context, a cluster of size $m=100$ corresponds to a diameter of approximately $4$nm, assuming hexagonal close packing and using $\bar{v} = 0.3$nm$^3$~\cite{heiss2010fetuin}. A diameter of $4$nm is similar in extent to the size of FA, which has been measured to have a hydrodynamical radius of $4.3$nm~\cite{heiss2003structural}.   We also note here that the radius of a PC is approximately $0.4$nm, so the ratio of the diffusivities between a PC monomer and FA molecule ($D/D_{\textrm{FA}}$), is approximately $10$.

To estimate the concentration of PC monomers ($\rho_\infty$), we rely on indirect evidence as precise quantification of these clusters does not appear to have been performed in the literature. We note that a study by  \citet{chughtai1968complexes} found that in physiological conditions approximately 6\% of  solution \Ca is present in calcium-phosphate complexes.  A separate study has found that approximately 50\% of calcium phosphate complexes have size consistent with Posner's cluster~\cite{oyane1999clustering}.  An ACP nucleation study using $2.5$mM free \Ca and $1$mM \ce{K2HPO4} found spherical clusters of approximately $30-80$nm in diameter after $1$hr~\cite{dey2010role}. Assuming that the $80$nm size corresponds to a cluster that nucleated soon after $t=0$, one finds that the concentration of PCs is at least $2$nM in their preparation.  Using this value, we estimate the an equilibrium constant for the formation of PCs
\begin{equation}
k_{\textrm{eq}} = \frac{\rho_\infty}{[\Ca]^3[\PO]^2}, \label{eq:keqpc}
\end{equation}
finding that $pk_\textrm{eq} \approx -16$. This compution estimates nanomolar-range concentrations for PCs in the physiological range that we defined above.

 With these rough estimates in mind, we may approximate an ``inhibition-ratio'' from Eq.~\ref{eq:inhibitionratio} as
\begin{equation}
I = \frac{m_p D_{\textrm{FA}}\phi_\infty }{D\rho_\infty} \approx \mathcal{O}(10^3).
\label{eq:inhibitionratio2}
\end{equation}
The concentration of post-nuclear clusters is exponential in this ratio, suggesting that large values for $I$ would inhibit calcification. We turn now to assays of calcification inhibition in order to validate this computation. 

\citet{heiss2008hierarchical} assessed the inhibition of sedimentation in a highly supersaturated solution of $20$mM \Ca and $6$mM \ce{Na2HPO4} in a closed system. For this system, Eq.~\ref{eq:keqpc} provides an estimate of $\rho_\infty = 2.5\times 10^{-5}$M. At $20\mu$M,  corresponding to $I=\mathcal{O}(10^2)$, FA was shown to inhibit sedimentation fully over a time interval of days. At $1.5\mu$M, corresponding to $I=\mathcal{O}(10^1)$, FA was seen to initially inhibit sedimentation, but only for a period of two hours. In the latter case, we expect two things to be occurring. First, because the experiments are conducted in closed systems, exhaustion of FA is occurring over the long time span. Second, inhibition is exponentially weaker than it is in physiological settings implying a proportionally quicker rate of sedimentation.

\section{Discussion}

\subsection{Protein consumption rate and implications}

The quantity $R$ sets a minimum replenishment rate for new FA protein in order to maintain a steady concentration
of FA, and hence colloidal stability.  As seen in Eq.~\ref{eq:bc}, the parameter $\lambda$ is present in the denominator of $c_0^\infty(s_*)$. As a result, to the leading order, $R$ increases as $\lambda$ decreases.
 Failure to maintain this replenishment rate leads to decrease in $\phi_\infty$, the concentration of FA. A drop in $\phi_\infty$ further decreases $\lambda$, thereby further exacerbating the situation (the less FA available, the more that is needed). Effectively, in the regime where calcium and phosphate concentrations remain supersaturated, the number of FA molecules required to buffer each nucleating mineral particle increases as the concentration of FA decreases.
 
  Thus, even a small destabilization in FA replenishment can feed-forward to avalanche into catastrophic calcium phosphate sedimentation. This observation explains the experimental finding that serum FA is often significantly depressed in systems exhibiting ectopic calcification, yet plentiful in the sedimented plaques~\cite{reynolds2005multifunctional}.
  
%  The knowledge that soft-tissue calcification is a prominent feature of many pathologies has caused health care providers to rethink the recommendation for calcium supplementation. Longitudinal studies have shown calcium supplementation to be associated with increased risk of heart disease~\cite{bolland2010effect} and stroke~\cite{li2012associations}. In addition, evidence for calcium supplementation's efficacy in preventing bone fracture is weak~\cite{bolland2015calcium}. Thermodynamically, the explanation is simple. As we have seen, calcium is normally available at high concentrations in fluids. So, its inclusion in bone is likely limited by other mechanisms as it is a highly-regulated process. Conversely, as shown in Fig.~\ref{fig:fig6}, increasing the calcium concentration increases both the nucleation rate for calcium phosphate as well as the growth rate of post-nuclear clusters. In particular, the consumption rate of FA is very sensitive to calcium concentration as it is in large-part driven by the nucleation rate. In turn, increased consumption of FA may drive FA concentration down, where a drastic change in sedimentation rate becomes apparent at approximately $5\mu$M in concentration.

\subsection{Assumptions, Limitations, and Extensions}

In our theoretical treatment of this topic, we have made some key simplifying assumptions. Correspondingly, we have  also limited the scope of our formulation and results.
We reiterate that we are primarily interested in the earliest stages of the mineralization process, immediately after nascent nuclei have overcome the kinetic barrier and progression is governed by thermodynamic considerations. For this reason, we do not consider later phases of calcium phosphates, as well as their nucleation through heterogeneous nucleation involving ACP precursors~\cite{jiang2015amorphous}.  It is notable, however, that the transformations of calcium phosphate from ACP to HA have been a rich topic of research, and FA protein is known to interact with these phases as well, just as it interacts with ACP~\cite{heiss2003structural}.

Biologically, we also assume that the specific structure of FA is important in two ways. First, its hydrodynamical mobility determines the rate at which it is able to interact with mineral clusters. Second, its precise biochemical structure determines its propensity for strong interactions with post-nuclear clusters. We have assumed this mechanism to be the primary mechanism for FA-based inhibition for several reasons. 

Mainly, this mechanism has been suggested in the experimental literature~\cite{rochette2009shielding}. Alternatively, FA has been reported to be able to bind calcium ions directly~\cite{suzuki1994calcium}, and hypothesized to bind pre-nucleation clusters (Posner's clusters) directly~\cite{heiss2010fetuin}. Yet, binding free calcium ions directly would not inhibit nucleation without decreasing supersaturation. Binding of pre-nucleation clusters would decrease the nucleation rate for calcium phosphates, running counter to experimental evidence~\cite{rochette2009shielding}.

Aside from the experimental evidence, there are apparent advantages to the solubilizing of post-nuclear particles as both an effective and efficient strategy for controlling sedimentation in super-saturated systems. While binding of pre-nucleation clusters would inhibit nucleation by increasing the energy barrier, it would require more inhibitory proteins to work because pre-nucleation clusters form at quicker rates than post-nucleation clusters. Experimental and theoretical evidence has confirmed the presence and stability of Posner's clusters in biological solutions. In our formulation, we assume that pre-nucleation clusters exist in quasi-steady equilibrium with free calcium and phosphate. We also assume that FA--free calcium and FA--pre-nucleation cluster interactions are weak and reversible so that we may ignore them.

It is this specificity of mechanism that makes FA a potent calcification inhibitor. Uniquely, experiments have shown that FA is the crucial protein for in-vivo mineralization inhibition. While other macromolecules such as albumin are known to interact with calcium phosphates, their physical attributes make them insufficient for this task.

  Albumin acts as a buffering agent for calcium in blood, helping to maintain \Ca concentration in an analogous manner to maintenance of free \ce{H+} ion concentration by pH buffers. 
For purposes of this study, the main effect of albumin is in setting the far-field equilibrium concentration of \Ca and hence mineral monomers. We are ignoring interactions between the regulatory FA protein and other plasma proteins such as albumin.  Although lacking in intrinsic capability, albumin has been shown to enhance the inhibitory properties of FA protein~\cite{pasch2012nanoparticle}, however, their main effect is in later-stage stabilization of complexes containing multiple protein-mineral clusters~\cite{heiss2008hierarchical}.

In this manuscript we have also ignored other possible contributing factors to the overall mineralization process including interactions with other ions such as sodium, chloride, magnesium, zinc, or H$^+$/OH$^-$. While these ions have been shown to influence mineralization, their importance in the early stages of nucleation is unclear. 

We also have not considered secondary interactions between mineral-FA hybrids, or the formation of calciprotein polymers. One of the goals of the present study has been to determine the content of the individual calciprotein monomers (the protein-mineral hybrid complexes we study in this manuscript). Observations by \citet{wald2011formation} have shown that the stability and size of these secondary structures varies with the concentration of FA present in the system. A possible cause for this effect is the variations in the mineral to protein ratio in the clusters that we form in our model. The understanding of these calciprotein monomers gained from this study should prove useful in better-understanding the kinetics behind the formation of calciprotein polymers as well as subsequent phase transitions.

Finally, while we have made an attempt to parameterize our model based on quantitative results from past literature, our numerical estimates remain rough guides and further experimentation is likely needed more-accurately parameterize our model. In particular, as the precise mechanism for calcium phosphate nucleation becomes better known, it will become easier to characterize the precise nucleation barrier and nucleation rate of calcium phosphate in order to better-understand the FA consumption rate. We have assumed for instance that the Posner cluster is the pre-nucleation cluster for this system, but recent literature has been mixed with respect to this hypothesis. We would like to emphasize, however, that our formulation is independent of the precise nature of these clusters as the bulk of our analyses focus on post-nucleation events.

Looking more broadly at our work, the methodology that we have developed in this manuscript has potential in explaining a variety of solubility problems throughout biology. As an example, the system of stabilization, transport, and clearance of lipid molecules by HDL and LDL bears striking resemblance to the calcium-phosphate-FA system that we have analyzed in this manuscript. The formation of protein-non-protein complexes or colloids is a widespread feature of the homeostasis of solutions in biology.

\section{Summary}
In this manuscript we have utilized classical nucleation theory to provide a quantitative description of the growth of calcium phosphate nanoparticles interacting with a shielding protein. In contrast with other theoretical work on similar systems, we have not neglected possible surface-limiting regimes of the process. Our quantitative description of the process provides an estimate of the critical concentration of shielding protein necessary for stable long-term inhibition of calcification, as well as an estimate of the total rate that the protein is consumed.

Critically, we have found that  disruptions of the ability to maintain the concentration of FA leads to \emph{increased} overall consumption of the protein, and hence, exhaustion and sedimentation.

\section{Acknowledgements}
The authors would like to thank  Lydia L. Shook (Yale School of Medicine),  Tom Chou (UCLA Mathematics and Biomathematics),
Huaxiong Huang (York University Mathematics), and Jonathan J. Wylie (City University of Hong Kong Mathematics)
for their comments and feedback relating to this work. This material is based upon work supported by the National Science
Foundation under Agreement No. 0635561. JC also acknowledges
support from the National Science Foundation through grant
DMS-1021818, and from the Army Research Office through grant 58386MA. This research was supported in part by the Intramural Research Program of the NIH, Clinical Center.

%\bibstyle{achemso}
\bibliographystyle{apsrev4-1}
\bibliography{cnp}

\end{document}